\documentstyle[epsfig,prb,aps,multicol]{revtex}
\begin{document}
\draft
\title{Electronic and dynamic properties of a one-dimensional
Thue-Morse chain}
\author{Gi-Yeong Oh}
\address{Department of Basic Science, Ansung National
 University, Kyonggi-do 456-749, Korea}
\date{\today}
\maketitle
\begin{abstract}
We study the electronic and dynamic properties of a one-dimensional
Thue-Morse chain within the framework of the transfer model. By means
of direct diagonalization of the Hamiltonian matrix we first show that
the trace map of the transfer matrices is exactly the same as that in
the diagonal model. Then, by calculating several quantities such as the
wave function, the Lyapunov exponent, and the Landauer resistivity, we
show that all the electronic states are extended despite the singular
continuity of the energy spectrum. Our results indicate that the
electronic properties of the Thue-Morse chain is independent of the
kind of the model, which is contrary to the result of Chakrabarti {\it
et al}. [Phys. Rev. Lett. {\bf 74}, 1403 (1995)]. To deepen our
understanding, we study the dynamics of an electronic wave packet and
show that the wave packet spreads superdiffusively for long times with
dynamic indices intermediate between periodic and quasiperiodic chains.
Several features of dynamics distinctive from the case of the Fibonacci
chain are discussed. We also study the effects of electron-phonon
interaction on the dynamics of the wave packet by taking into account a
kind of nonlinear interaction. Degree of dynamic localization is shown
to be crucially dependent on the strength of the hopping energy and the
nonlinear interaction parameter.
\end{abstract}
\pacs{PACS numbers: 71.25.-s, 61.44.+p, 64.60.Ak, 71.20.Ad}

\begin{multicols}{2}


\section{Introduction}

Since the discovery of the quasicrystalline phase$^{1}$ much attention
has been devoted to the quasiperiodic systems. The Fibonacci chain, a
one-dimensional version of quasicrystal characterized by either the
inflation rule $(L,S)\rightarrow (LS,L)$ or the substitution rule
$S_{l+1}=S_{l}S_{l-1}$ with $S_{1}=\{ L\}$, is the most intensively
studied example, and new concepts on the electronic properties such as
the singular continuous energy spectrum and the critical states are now
well established.$^{2}$ Stimulated by unusual electronic properties of
the Fibonacci chain as well as the development of layer-growth
technique,$^{3}$ there has been a variety of interest in
one-dimensional deterministic aperiodic chains. Among them the most
well-known example is the Thue-Morse chain characterized by either the
inflation rule $(L,S)\rightarrow (LS,SL)$ or the substitution rule
$S_{l+1}=S_{l}\overline{S}_{l}$ with $S_{1}=\{L\}$.

Unusual electronic and Fourier spectral properties of the Thue-Morse
chain are often compared with those of the Fibonacci chain. According
to the nature of the Fourier spectrum of the sequence, the Thue-Morse
sequence is more disordered than the Fibonacci sequence since the
former has a singular continuous spectrum$^{4}$ while the latter has a
point-like spectrum with Bragg $\delta$-peaks. In the meanwhile,
according to the nature of the electronic energy spectrum, both of the
chains are located in the same category; both have the singular
continuous energy spectrum, which is a general feature of
one-dimensional deterministic aperiodic chains.$^{5}$ However,
according to the nature of the electronic states, the Thue-Morse chain
is much closer to the periodic chain than the Fibonacci chain; all the
electronic states of the latter are rigorously proven to be
critical,$^2$ while most of the electronic states of the former exhibit
Bloch-like extended character.$^{6}$

In studying the electronic properties of one-dimensional deterministic
aperiodic chains, it has been common to use the tight-binding equation
\begin{equation}
 t_{n+1,n}\psi_{n+1}+t_{n-1,n}\psi_{n-1}+v_{n}\psi_{n}=E\psi_{n}
\end{equation}
or
\begin{equation}
\left(\begin{array}{c}\psi_{n+1}\\ \psi_{n}\end{array}\right)
=T_{n}\left(\begin{array}{c}
\psi_{n}\\ \psi_{n-1}\end{array}\right).
\end{equation}
Here $v_{n}$ is the on-site energy of the site $n$, $t_{n\pm 1, n}$
is the nearest neighboring hopping energy between the sites $n$ and
$n\pm1$, and $T_{n}$ is the transfer matrix.

When a sequence to be studied is introduced into $v_{n}$ ($t_{nm}$)
with setting $t_{nm}=1$ ($v_{n}=0$), it is called the diagonal
(transfer) model. In the case of the Fibonacci chain, there is no
preference in choosing the kind of the model and both of the models
have been equally well studied. However, in the case of the Thue-Morse
chain, most of the existing literatures have been devoted to the
diagonal model rather than the transfer model because the transfer
matrix $M_{l}=\Pi_{n=1}^{N}T_{n}$ appeared in the transfer model is so
complex that the trace map method which is the most powerful tool in
studying quasiperiodic or aperiodic chains cannot be easily
applied.$^{7}$ Furthermore, the complexity of the transfer matrix in
the transfer model also leads a literature that contains incorrect
argument on the electronic properties of the Thue-Morse chain:
Chakrabarti {\it et al}.$^{8}$ argued that the transfer model does not
support the extended states unlike to the case of the diagonal model.
However, we argue that the electronic properties of the Thue-Morse
chain are independent of the kind of the model. To certify our
argument, we show that the trace map of the transfer matrices in the
transfer model is exactly the same as that in the diagonal model and
that all the electronic states except the two edge states are extended.

It has been generally believed that the electronic states corresponding
to a singular continuous energy spectrum are critical. Of course, in
some cases such as the generalized Fibonacci chain$^{9}$ and the
period-doubling chain,$^{10}$ existence of extended states has been
reported. However, even in these cases, the portion of the extended
states to the critical states goes to zero in the infinite limit of the
chain size, and the global electronic properties of the chains are
qualitatively similar to those of the Fibonacci chain. In this context,
the Thue-Morse chain is very exceptional since most of the electronic
states exhibit extended nature despite of the singular continuity of
the energy spectrum,$^{5}$ and it may be interesting to investigate the
phenomena resulting from these peculiar electronic properties. By the
way, one of the interesting properties of the chains with a singular
continuous energy spectrum lies in the quantum dynamics of an
electronic wave packet. The spread of the wave packet in the Fibonacci
chain is known to be an anomalous power-law diffusion; the spread can
be diffusive, subdiffusive, or superdiffusive, depending on the
strength of the system parameters.$^{11}$ Since the nature of the
energy spectrum is known to play an important role in quantum dynamics
of the wave packet,$^{12}$ a chain with the singular continuous
spectrum is expected to exhibit similar dynamic property to that of the
Fibonacci chain. Then, it may be worthwhile to test whether the dynamic
property of the Thue-Morse chain is similar to that of the Fibonacci
chain or not. We try to answer this question and discuss several
characteristics distinctive from the case of the Fibonacci chain.

Since Eq.~(1) is the Schr\"{o}dinger equation for a single electron,
effects of electron-electron or electron-phonon interactions, that are
inevitable in real systems, are neglected. Thus it may be questionable
whether the single-electron properties obtained from Eq.~(1) can be
observable or not when these interactions are present. As a step to
solve this question, we take into account a kind of nonlinear
interaction and study the effect of the interaction on the dynamics of
the electronic wave packet. We show that degree of dynamic localization
is crucially dependent on the strength of the hopping energy and the
interaction parameter.

This paper is organized as follows: In Sec. II we deduce the trace map
of the transfer matrices in the transfer model by directly
diagonalizing the Hamiltonian matrix of the Thue-Morse chain. Then we
elucidate the nature of the electronic states by calculating the wave
functions, the Lyapunov exponents, and the Landauer resistivities of
the states. In Sec.~III we illustrate the dynamic properties of an
electronic wave packet by observing the evolution of the time-dependent
Schr\"{o}dinger equation. And, in Sec.~IV we present the effect of the
nonlinear interaction on the dynamics of the wave packet. Finally a
summary is given in Sec.~V.


\section{Electronic Properties of the Thue-Morse chain}

As a first step to examine the electronic properties in the framework
of the transfer model, we deduce the trace map of the transfer
matrices. To do this, we first consider the characteristics of the
energy eigenvalues obtained by directly diagonalizing an $N\times N$
Hamiltonian matrix under a periodic boundary condition. Figure~1 shows
a plot of $E$ versus $l$, where $l$ is the order of the chain size
($N=2^{l}$) and $T (\equiv t_{L}/t_{S})$ is the strength of the hopping
energy. From this, we obtain the following characteristics: (a) The
eigenvalues are symmetric with respect to $E=0$ and there always exists
a state with $E=0$. (b) All the eigenvalues except the two outermost
eigenvalues are doubly degenerate, and there are $(2^{l-1}+1)$
distinctive eigenvalues at the $l$th-order Thue-Morse chain. (c) For
$l\ge 2$, all the eigenvalues at the $l$th-order chain remain as those
at the $(l+1)$th-order chain. (d) Additional eigenvalues at the
$(l+1)$th-order chain are located between the eigenvalues at the
$l$th-order chain.

The trace map of the transfer matrices can be deduced as follows:
First, by noting that property (c) implies the fact that $(x_{l-1}-1)$
is an ingredient of $(x_{l+1}-1)$, we can write
$x_{l+1}-1=f(l)(x_{l}-1)$, where $x_{l}$ is the half of the trace of
$M_{l}$. Second, by means of properties (c) and (d), we draw the
eigenvalue tree, as in Fig.~2. Since the tree resembles that of the
diagonal model,$^{6}$ we assume that $f(l)$ is a power of $x_{l-1}$. In
order to check this assumption, we calculate the energies
$\{E_{l-1}^{'}\}$ satisfying $x_{l-1}=0$ to compare them with the
energies $\{E_{l+1}\}$ satisfying $x_{l+1}=1$. Then, we find that
$\{E_{l-1}^{'}\}$ becomes a half part of $\{E_{l+1}\}$, i.e.,
$\{E_{l+1}\}=\{E_{l-1}^{'} ,E_{l}\}$. Thirdly, by combining this result
with property (b), we find that $f(l)$ is a square function of
$x_{l-1}$, i.e, $f(l)=4x_{l-1}^{2}$. Thus, the trace map of the
transfer matrices becomes
\begin{equation}
x_{l+1}=4x_{l-1}^{2}(x_{l}-1)+1
\end{equation}
with $x_{1}=[E^{2}-(t_{L}^{2}+t_{S}^{2})]/2t_{L}t_{S}$ and
$x_{2}=E^{2}[E^{2}-2(t_{L}^{2}+t_{S}^{2})]/2t_{L}^{2}t_{S}^{2}+1$.

Using Eq.~(3), we can calculate energy eigenvalues for any chain size.
For example, let us consider the case of $T=0.5$. Then, since we can
easily obtain the energies $E_{1}^{'}=\{\pm\sqrt{5}\}$,
$E_{2}^{'}=\{\pm\sqrt{5\pm\sqrt{17}}\}$, $E_{2}=\{0,\pm\sqrt{10}\}$
and $E_{3}^{'}=\{\pm[5\pm[(25\pm\sqrt{497})/2]^{1/2}]^{1/2}\}$,
we can write the energy eigenvalues up to $l=5$ exactly:
$E_{3}=\{0,~\pm{\sqrt{5}},~\pm{\sqrt{10}} \}$,
$E_{4}=\{0,~\pm{\sqrt{5}},~\pm{\sqrt{10}},~\pm\sqrt{5\pm\sqrt{17}}\}$,
and $E_{5}=\{0,~\pm{\sqrt{5}},~\pm{\sqrt{10},~\pm
\sqrt{5\pm\sqrt{17}},~\pm[5\pm[(25\pm\sqrt{497})/2]^{1/2}]^{1/2}}\}$.
We have checked that, for higher $l$, the energy eigenvalues obtained
numerically by Eq.~(3) are exactly the same as those obtained from
direct diagonalization method within the numerical accuracy.

Before going on further, we would like to mention some remarks. The
first is that Eq.~(3) is the same form as that of the diagonal model.
This is a very surprising result because the transfer model has a much
more complex pattern of the transfer matrices compared with those of
the diagonal model. The second is that we deduced Eq.~(3) by means of a
numerical calculation. Thus, the problem of a mathematically rigorous
derivation of Eq.~(3) still remains as an open problem. However, our
purpose in this paper is not to solve this problem exactly, but to
elucidate the nature of the electronic energy eigenvalues and the
eigenstates, let us remain it open. The third is that, by using the
branching rule, the number of any desired subband can be indexed in the
same way as in Ref.~6. This means that one can also perform a scaling
analysis on the energy bandwidths as in the case of the diagonal model,
which is in contrast to the argument of Ref.~8.


From now on, we discuss the nature of the eigenstates with energies
obtained by Eq.~(3). To this end, let us first classify the eigenstates
into two groups - the degenerate states and the edge states. Since
Eq.~(3) is exactly the same as that of the diagonal model except for
the initial conditions $x_{1}$ and $x_{2}$, we affirm that the same
behavior as that of the diagonal model also appears in the transfer
model; degenerate eigenstates at the $l$th-order chain will exhibit an
$m$th-order-type chain-like extended behavior in the wave functions at
the $(l+m)$th-order chain. Numerical calculations verify our
affirmation. We present some examples in Fig.~3. The only exceptions
are the edge states with energies satisfying $x_{2}=1$; the wave
functions diverge algebraically with increasing the chain size, as in
the edge states of the periodic chain.


A little cumbersome calculation enables us to write $M_{l}$ as
\begin{equation}
 M_{l}=I
\end{equation}
for the degenerate states and
\begin{equation}
 M_{l}=\left(\begin{array}{cc} 1+aN & bN \\
 cN & 1-aN \end{array}\right)
\end{equation}
for the edge states. Here, we have set $a=r^{2}/4$,
$b=\mp(r^{2}/8t_{L})\sqrt{2(t_{L}^{2}+t_{S}^{2})}$,
$c=\mp(2t_{L}/rt_{S})b$, and $r\equiv t_{L}/t_{S}+t_{S}/t_{L}$.
Using Eqs.~(4) and (5), we can obtain the Lyapunov exponent $\zeta$
defined by
\begin{equation}
 \zeta\equiv\lim_{N\rightarrow\infty}\zeta_{N}
 =\lim_{N\rightarrow\infty}\frac{1}{N}\ln||M_{l}||,
\end{equation}
where $||~||$ is the modulus of the matrix. Thus, the localization
length $\xi_{N}~(=\zeta_{N}^{-1})$ for a chain size $N$ is given by
$\xi_{N}=N/\ln2$ for the degenerate states and $\xi_{N}\approx N/ 2\ln
N$ for the edge states; $\xi_{N}$ goes to infinity in the infinite
limit of the chain size, which indicates the absence of exponentially
localized states.


As another test of the nature of the eigenstates, we calculate the
Landauer resistivity. Assuming that a Thue-Morse chain of size $N$ is
embedded in the middle of a perfect conductor and that an electron
with $E=2T\cos{k}$ comes in from the left-hand side of the perfect
conductor, the dimensionless resistivity can be written as
\begin{equation}
\rho(k,N)=\frac{1}{N}\left|U_{21}(k,N)\right|^{2},
\end{equation}
where $U_{21}(k,N)$ is the $(2,1)$ component of the transfer matrix
relating the coefficients of the incoming and reflecting plane waves
to that of the transmitted plane wave. [See Ref.~13 for the detailed
formalism.] Using Eqs.~(4) and (5), we obtain that
$|U_{21}(k_{0},N)|^{2}=\cot^{2}k_{0}$ for even $l$ and
$|U_{21}(k_{0},N)|^{2}=(t_{S}/t_{L})^{2}\cot^{2}k_{0}$ for odd $l$,
while $|U_{21}(k_{1},N)|^{2}\sim N^{2}$. Here, $k_{0}$ and $k_{1}$
are the wave numbers that correspond to the degenerate and the edge
eigenvalues. Thus, we have
\begin{equation}
 \rho(k,N)\sim\
\left\{\begin{array}{ccc} 1/N&\makebox{for} & k=k_{0} \\
 N&\makebox{for} & k=k_{1} \end{array}\right. ,
\end{equation}
which indicates that the degenerate (edge) states are extended
(algebraically localized).


\section{Dynamic Properties of the Thue-Morse chain}


To study the dynamic properties of the Thue-Morse chain, we consider
the time--dependent Schr\"{o}dinger equation
\begin{equation}
 i\frac{\partial\psi_{n}}{\partial t}=H_{0}\psi_{n}
\end{equation}
where
\begin{equation}
 H_{0}\psi_{n}=t_{n+1,n}\psi_{n+1}+t_{n-1,n}\psi_{n-1}+v_{n}\psi_{n}.
\end{equation}
In integrating Eq.~(9), we assume an electronic wave packet locates
initially at $\psi_{n}=\delta_{n,n_{0}}$ and observe the time evolution
of the wave packet by employing the fourth-order Runge-Kutta algorithm.
Besides, we take the chain size as $N=16384$ (i.e., $l=14$) with
$n_{0}=8192$ and the time step as $\Delta t=0.05$. Note that, even
though we use the finite chain size and the fixed boundary condition
($\psi_{0}=\psi_{N}=0$), they have no effect on the results of the
integration since we perform the integration within time interval where
the wave front does not reach the boundary region. We also check the
accuracy of the numerical integration by monitoring the conservation of
the total probability (i.e., $\sum_{n} |\psi_{n}|^{2}=1$).

In clarifying the dynamic nature of the wave packet, the most important
quantity is the variance of the wave packet
\begin{equation}
 V(t)=\sum_{n}(n-n_{0})^{2}|\psi_{n}(t)|^{2},
\end{equation}
which gives a global estimate of the spread of the wave packet in
space. The asymptotic long-time behavior of the variance follows the
power-law
\begin{equation}
 V(t)\sim t^{\gamma}.
\end{equation}
It is well known that $\gamma=0$ for the localization, $0<\gamma<1$
for the subdiffusion, $\gamma=1$ for the ordinary diffusion,
$1<\gamma<2$ for the superdiffusion, and $\gamma=2$ for the ballistic
motion, respectively.

Figures~4 and 5 show the results of $V(t)$ for several values of the
system parameters $T$ and $v$ ($\equiv v_{L}=-v_{S}$). The curves in
Fig.~4 (5) are obtained within the framework of the transfer (diagonal)
model. For the curves in Fig.~4, we obtain the dynamic index $\gamma$
as $\gamma=1.942\pm 0.394$ for $T=0.125$, $\gamma=1.931\pm 0.103$ for
$T=0.2$, $\gamma=1.924\pm 0.039$ for $T=0.5$, $\gamma=1.998\pm 0.001$
for $T=1.0$, $\gamma=1.853\pm 0.028$ for $T=1.25$, $\gamma=1.924\pm
0.044$ for $T=2.0$, and $\gamma=1.840\pm 0.078$ for $T=5.0$,
respectively. Note that the standard deviation in the calculated
$\gamma$'s becomes larger as the value of $T$ deviates from $T=1.0$,
which means that the oscillatory behavior becomes stronger as
$T\rightarrow\infty$ and $T\rightarrow 0$. However, from the trend of
the curves, we expect that the oscillatory behavior will shrink to zero
for sufficiently long times. Concerning with the oscillatory behavior,
we would like to mention a point: Comparing Fig. 4 (5) with Fig.~2 (6)
of Ref.~11, we can see that there is no strong hierarchical time
evolution of $V(t)$ unlike to the case of the Fibonacci chain. This
feature may attribute to the weak self-similarity in the energy
spectrum of the Thue-Morse chain.

Recently, de Brito {\it et al.}$^{14}$ argued with the diagonal model
of the Thue-Morse chain that $\gamma$ is independent of the strength of
$v$ and is given by $\gamma=1.65$. On the contrary, Katsanos {\it et
al.}$^{15}$ argued with the same model that $\gamma$ is strongly
dependent on the strength of $v$ and given by $\gamma=1.5$ for large
values of $v$. However, judging from our results, both of the arguments
are partly correct. As can be seen in Figs.~4 and 5, $\gamma$ depends
on the system parameters $T$ and $v$ to a certain extent unlike to the
result of Ref.~14. However, the dependence seems not to be strong
unlike to the argument of Ref.~15. Note that this is another feature of
the Thue-Morse chain distinctive from the Fibonacci chain: In the case
of the Fibonacci chain, $\gamma$ decreases continuously from $2$ to $0$
as the system parameters increase, and the diffusion of the wave packet
is known to be either superdiffusive or subdiffusive, depending upon
whether $T^{-1}\leq 4$ or $T^{-1}\geq 4$.$^{11}$ However, there seems
to be no such critical value of $T$ in the Thue-Morse chain and the
spread of the wave packet seems to be always superdiffusive;
$1.8\le\gamma<2.0$ for $T>1$ and $1.9\le\gamma<2.0$ for $T<1$ in the
transfer model and $\gamma\simeq 1.50\sim 1.65$ for the diagonal model.
The most important point to be noted is the fact that the obtained
$\gamma$'s are intermediate between the value in the perfect periodic
chain ($\gamma=2$) and the value in the metallic regime of the Harper's
model or three-dimensional disordered lattices ($\gamma=1$). This
feature attributes to the unusual electronic properties of the
Thue-Morse chain; $\gamma>1$ reflects the extended nature of the
electronic states, while $\gamma<2$ reflects the singular continuity of
the energy spectrum.

Another important quantity that characterizes the dynamics of the wave
packet is the initial-site probability defined by
$R(t)=|\psi_{n_{0}}(t)|^{2}$, which has been widely used in
investigating the self-trapping transition of the wave packet.$^{16}$
However, since the fluctuation in $R(t)$ is too large to extract
characteristic features of the dynamics, we calculate the temporal
autocorrelation function$^{17}$
\begin{equation}
 C(t)=\frac{1}{t}\int_{0}^{t}R(t^{'})dt^{'},
\end{equation}
which is the time-averaged quantity of the initial-site probability.
The decay of $C(t)$ for long times is known to follow the power-law
\begin{equation}
 C(t)\sim t^{-\delta}.
\end{equation}
The exponent $\delta$ approaches $1.0$ for long times in the periodic
chain, while it goes to $0$ in the disordered chain. Figure~6 shows
the results of $C(t)$ for several values of $T$ in the transfer model,
where the power-law decaying behavior of $C(t)$ is clearly seen;
$\delta\sim 0.897$ for $T=1.0$, $\delta\sim 0.313$ for $T=1.25$,
$\delta\sim 0.214$ for $T=2.0$, and $\delta\sim 0.114$ for $T=5.0$,
respectively. The exponent $\delta$ decreases with increasing $T$,
which implies that the spread of wave packet becomes slower as $T$
increases.

We also calculate the participation number
\begin{equation}
 P(t)=\left\{\sum_{n}|\psi_{n}(t)|^{4}\right\}^{-1},
\end{equation}
which gives a rough estimate of the number of sites where the wave
packet has a significant amplitude; $P(t)=1$ for a single-localized
state and $P(t)=N$ for a state uniformly extended over $N$ sites.
Figure~7 shows the results of $P(t)$ for several values of $T$ in the
transfer model. $P(t)$ increases almost monotonically for $T=1.0$,
while it increases with strong oscillation for $T>1.0$. The oscillation
in $P(t)$ increases and the values of $P(t)$ rapidly decreases with
increasing $T$. The former reflects the singular continuity of the
energy spectrum and the latter implies that it becomes more difficult
for a wave packet to propagate.


\section{Nonlinear Interaction and the Dynamics}


To study the effect of the electron-phonon interaction on the dynamics
of the wave packet we take into account a kind of nonlinear
interaction$^{18}$ and consider the time--dependent Schr\"{o}dinger
equation
\begin{equation}
 i\frac{\partial\psi_{n}}{\partial t}=(H_{0}+H_{1})\psi_{n}
\end{equation}
where
\begin{equation}
 H_{1}=-\alpha|\psi_{n}|^{2}\psi_{n}.
\end{equation}
Here $\alpha$ gives the strength of the nonlinear interaction. Note
that the nonlinear interaction we consider describes a static
short-range electron-phonon interaction, which appears in many
situations such as, for example, the Holstein model for polaron
theory.$^{19}$

Dynamics of an electronic wave packet is known to be sensitively
dependent on the nonlinear term in Eq.~(16). In the case of periodic
chains,$^{20}$ there occurs self-trapping transition when $\alpha$
becomes larger than a certain critical value $\alpha_{c}$. And, in
disordered chains,$^{21}$ degree of dynamic localization weakens with
the increase of $\alpha$. In the meanwhile, in the case of the
Thue-Morse chain, Johansson {\it et al}.$^{12}$ argued with the
diagonal model that self-trapping occurs for arbitrarily small values
of $\alpha$ and that the variance of the wave packet increase
infinitely with time. Thus it may be questionable whether the dynamics
of the wave packet within the transfer model is different from the case
of the diagonal model or not. We integrate numerically Eq.~(16) with
setting $v=0$.

Figure~8 shows the variance of the wave packet for several values of
$\alpha$. As in the case of the linear case ($\alpha=0$), the
oscillatory behavior shrinks to zero for long times. Thus the
asymptotic long-time behavior of the wave packet is also characterized
by a superdiffusive movement in spite of the existence of the
nonlinearity, which resembles the cases of the linear transfer model
and the nonlinear diagonal model.$^{12}$ Besides, the dynamic index
$\gamma$ is nearly independent of the strength of $\alpha$, which also
resembles the case of the nonlinear diagonal model. The most
interesting point to be noted is that the dependence of the variance on
the strength of $\alpha$ for $T>1$ is qualitatively different from that
for $T<1$: When $T>1$, the variance decreases with increasing $\alpha$
[Fig.~8(a)], which indicates that the nonlinearity resists the
propagation of the wave packet. In the meanwhile, when $T<1$, the
variance increases with increasing $\alpha$ [Fig.~8(b)], which
indicates that the nonlinearity assists the propagation of the
electronic wave packet. Our results show that degree of dynamic
localization of the wave packet in the Thue-Morse chain crucially
depends on the strength of the interaction.


\section{Summary}

In summary, we study the spectral and dynamic properties of a
one-dimensional Thue-Morse chain mainly within the framework of the
transfer model. By using the method of direct diagonalization of the
Hamiltonian matrix, we first illustrate the energy spectral properties
and deduce the trace map of the transfer matrices. The trace map is
shown to be exactly the same as that in the diagonal model. Then we
illustrate the nature of the electronic states by calculating the wave
functions, the Lyapunov exponents, and the resistivities of the states.
All the doubly degenerate eigenstates are shown to be extended, while
the edge states are algebraically localized. Next, in the study of the
dynamics, we show that the asymptotic long-time behavior of an
electronic wave packet is superdiffusive with the dynamic indices
$\gamma$ satisfying $1<\gamma<2$; $\gamma>1$ reflects the extended
nature of the electronic states, while $\gamma<2$ reflects the singular
continuity of the energy spectrum. We also show that the dependence of
the dynamic index on the system parameters is much weaker than that of
the Fibonacci chain and that there is no hierarchical time evolution of
the variance unlike to the Fibonacci chain. Finally, by taking into
account a kind of nonlinear interaction, we study the effect of the
electron-phonon interaction on the dynamics of the wave packet and show
that degree of the localization of the wave packet crucially depends on
the strength of the interaction.

In spite of fundamental importance of electron-electron interaction, we
did not take into account the interaction in this paper. However, note
that, in connection with the problem of whether degree of localization
of a single-electron weakens or strengthens in the presence of this
interaction, much attention has been devoted to the chains with
disordered$^{22}$ and quasiperiodic$^{23}$ potential energies. Thus, it
is worthwhile to investigate the effect of electron-electron
interaction on the electronic and dynamic properties of the Thue-Morse
chain. This problem is on working and will be published elsewhere.

\vskip 0.3in
\centerline{\bf ACKNOWLEDGMENTS}
\vskip 0.2in
This work was financially supported by the Korea Research Foundation
made in the program of 1997-003-D00092.


\vskip 0.2in
\centerline{\large\bf FIGURE CAPTIONS}
\vskip 0.2in

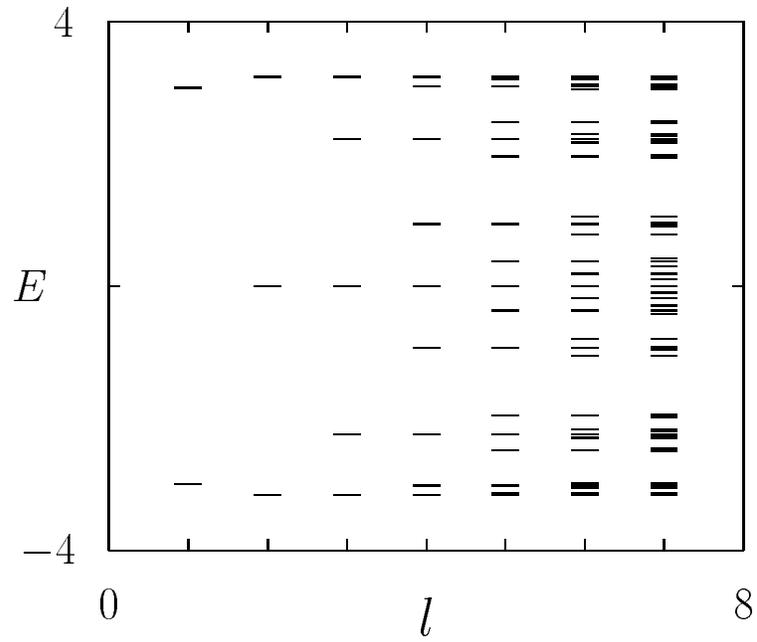
\noindent{\bf Fig.~1}.
$E$ versus $l$ in the transfer model with $T=0.5$.

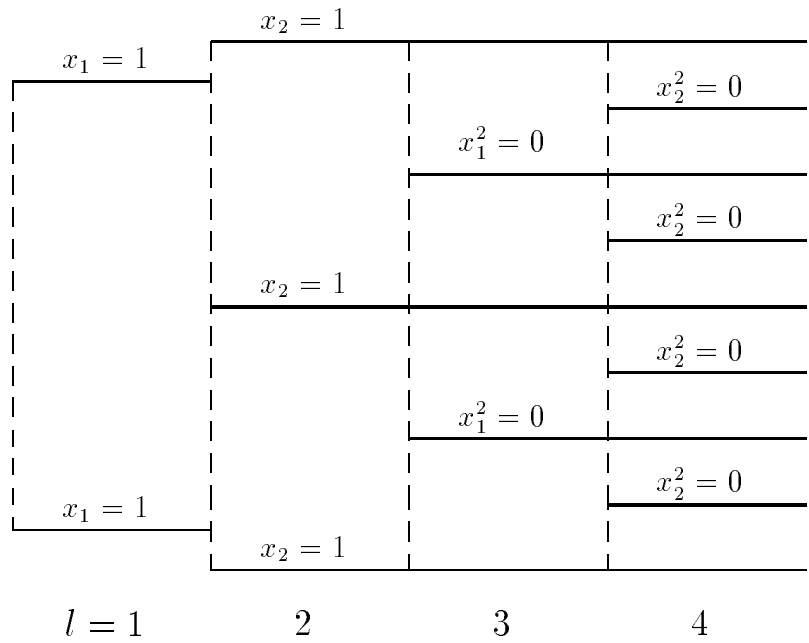
\noindent{\bf Fig.~2}.
Schematic representation of the eigenvalue tree.

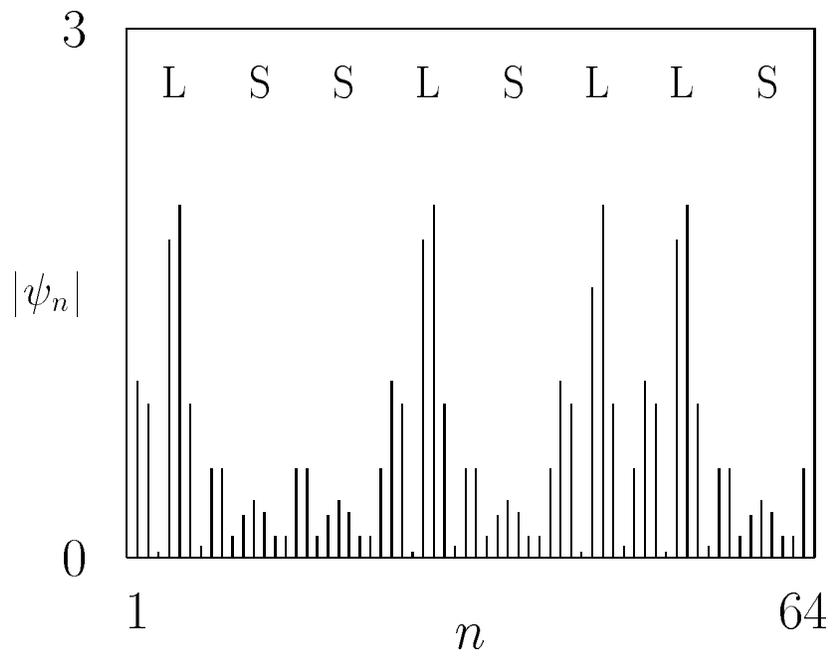
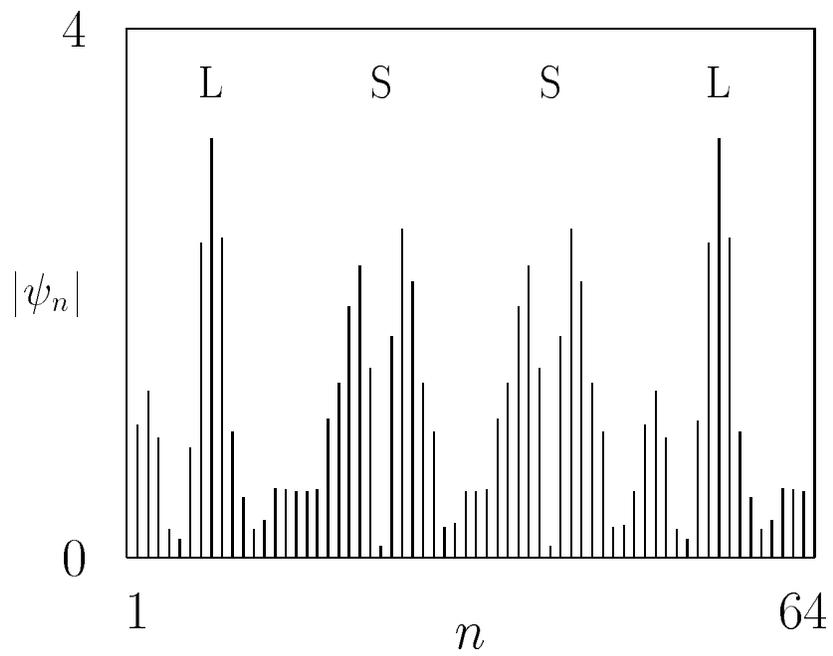
\noindent{\bf Fig.~3}.
$|\psi_{n}|$ versus $n$ for states with (a) $E=\sqrt{5}$
($x_{1}=0$) and (b) $E=\sqrt{5+\sqrt{17}}$ ($x_{2}=0$) with $T=0.5$,
$(\psi_{0},\psi_{1})=(0.5,1)$, and $l=6$. $|\psi_{n}|$ exhibits a
third-order-type (i.e., $LSSLSLLS$-type) chain-like behavior for (a)
and a second-order-type (i.e., $LSSL$-type) chain-like behavior for
(b).

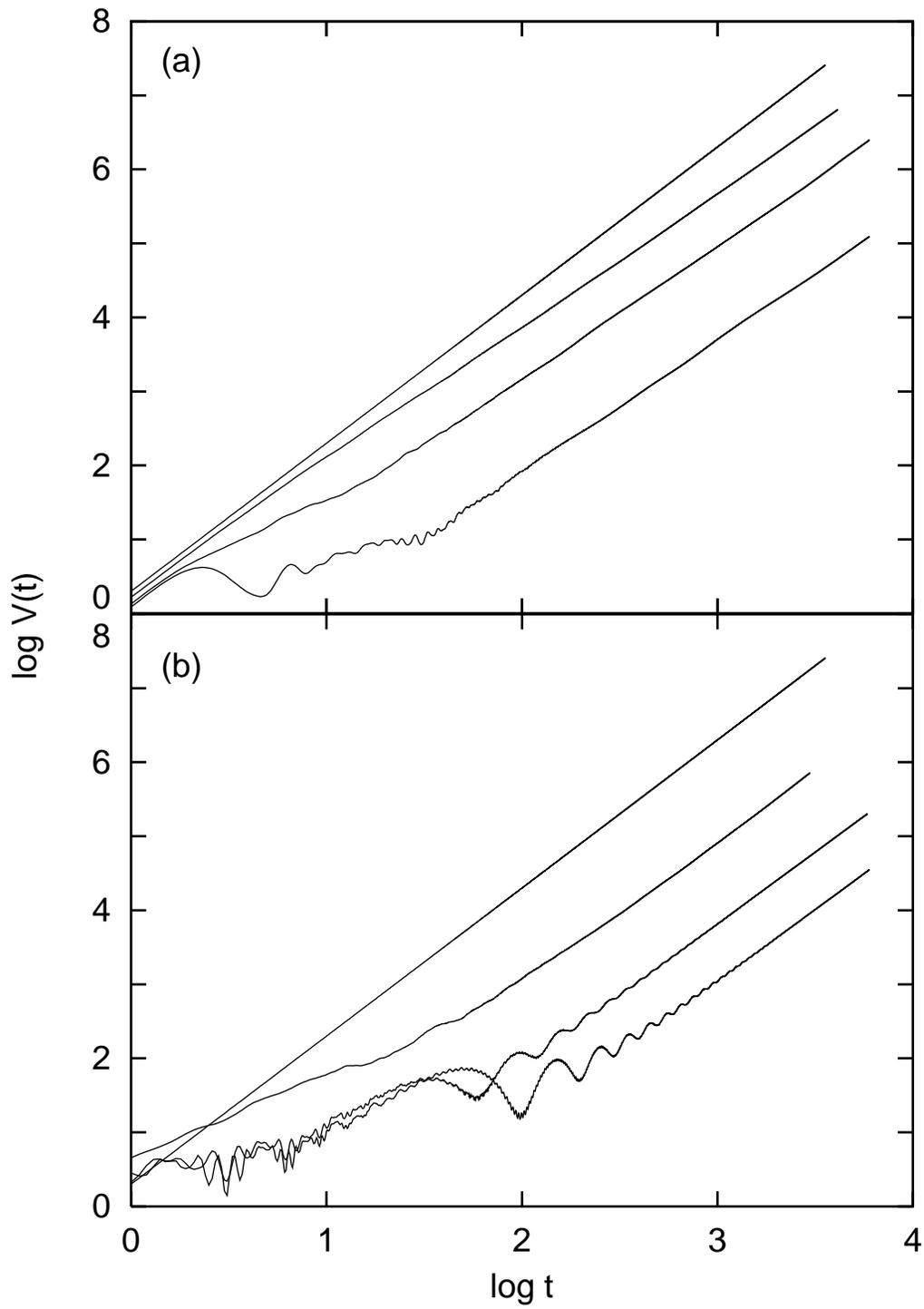
\noindent{\bf Fig.~4}.
$\log_{10}V(t)$ versus $\log_{10}t$ within the transfer model: (a)
$T=1.0, 1.25, 2.0, 3.0$, and $5.0$, and (b) $T=1.0, 0.5, 0.2$, and
$0.125$ from top to bottom at $t=1000$.

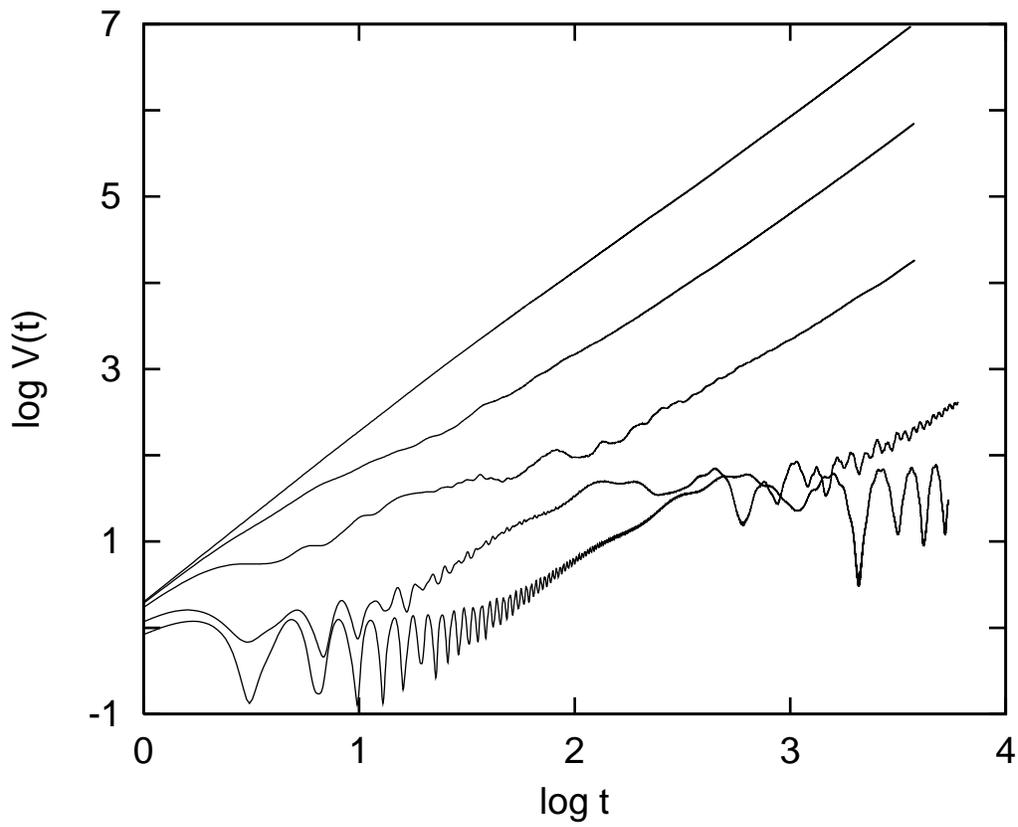
\noindent{\bf Fig.~5}.
$\log_{10}V(t)$ versus $\log_{10}t$ within the diagonal model, where
$v=0.1, 0.5, 1.0, 2.0$, and $3.0$ from top to bottom at $t=3000$.

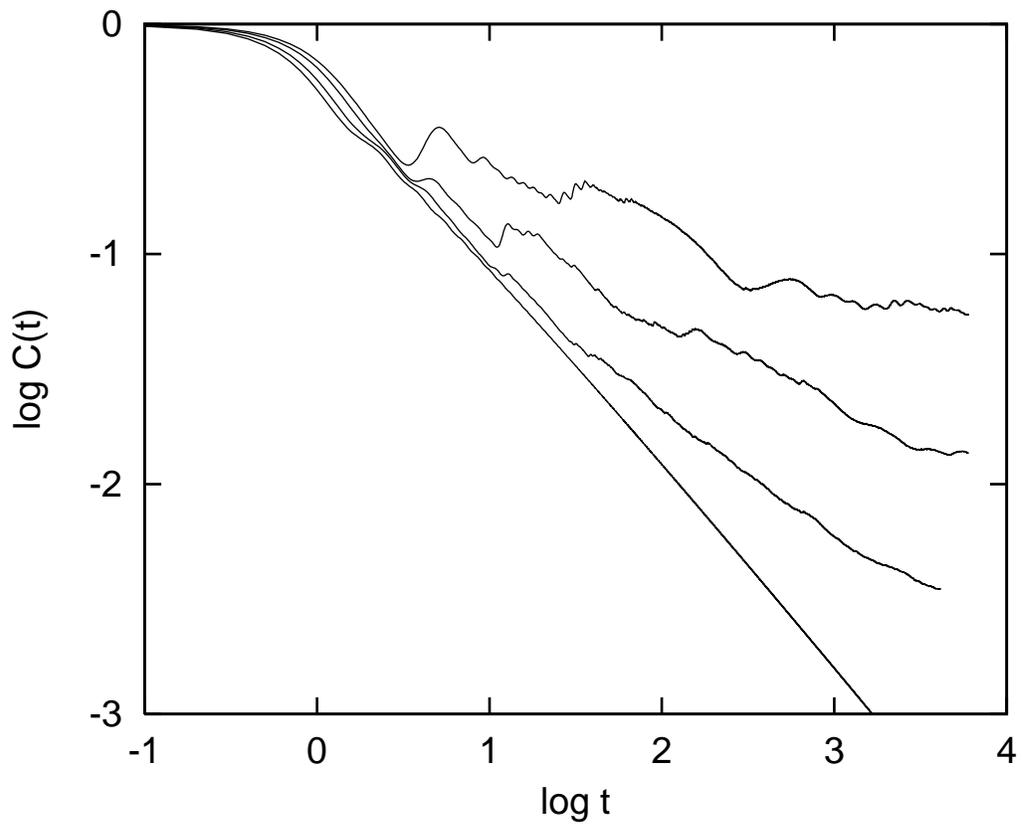
\noindent{\bf Fig.~6}.
$\log_{10}C(t)$ versus $\log_{10}t$ within the transfer model, where
$T=1.0, 1.25, 2.0$, and $5.0$ from bottom to top at $t=1000$.

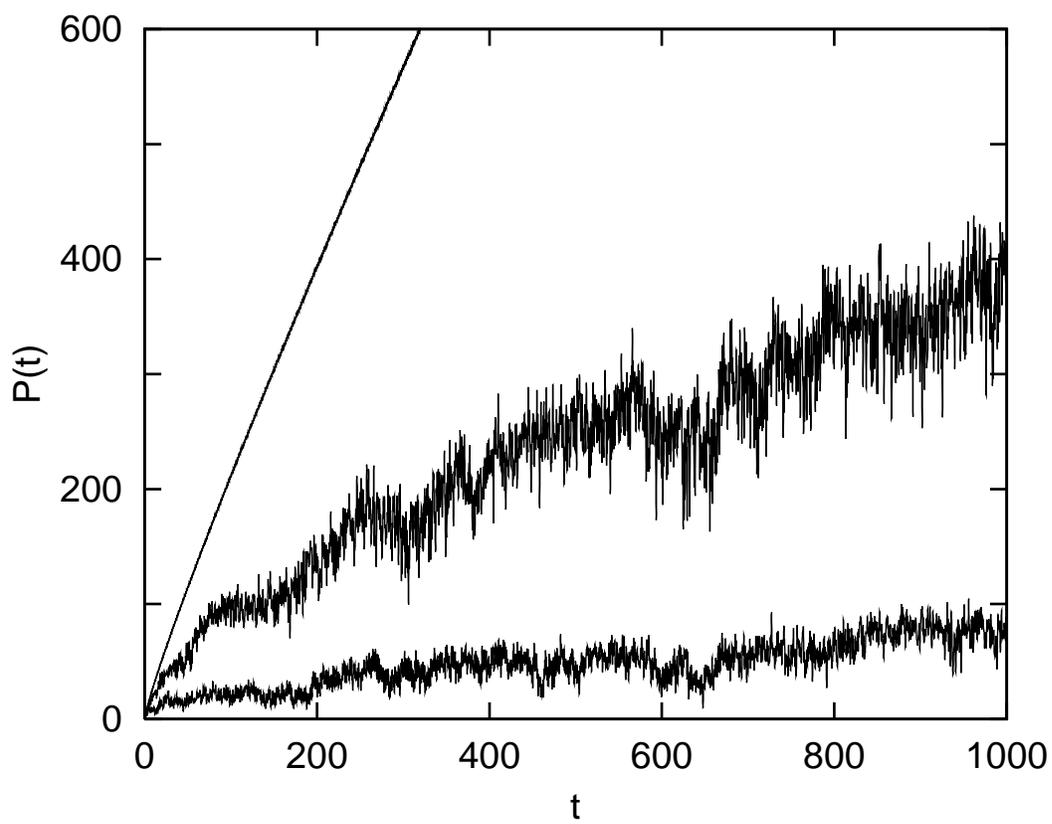
\noindent{\bf Fig.~7}.
$P(t)$ versus $t$ within the transfer model, where $T=1.0, 1.25$, and
$2.0$ from top to bottom at $t=200$.

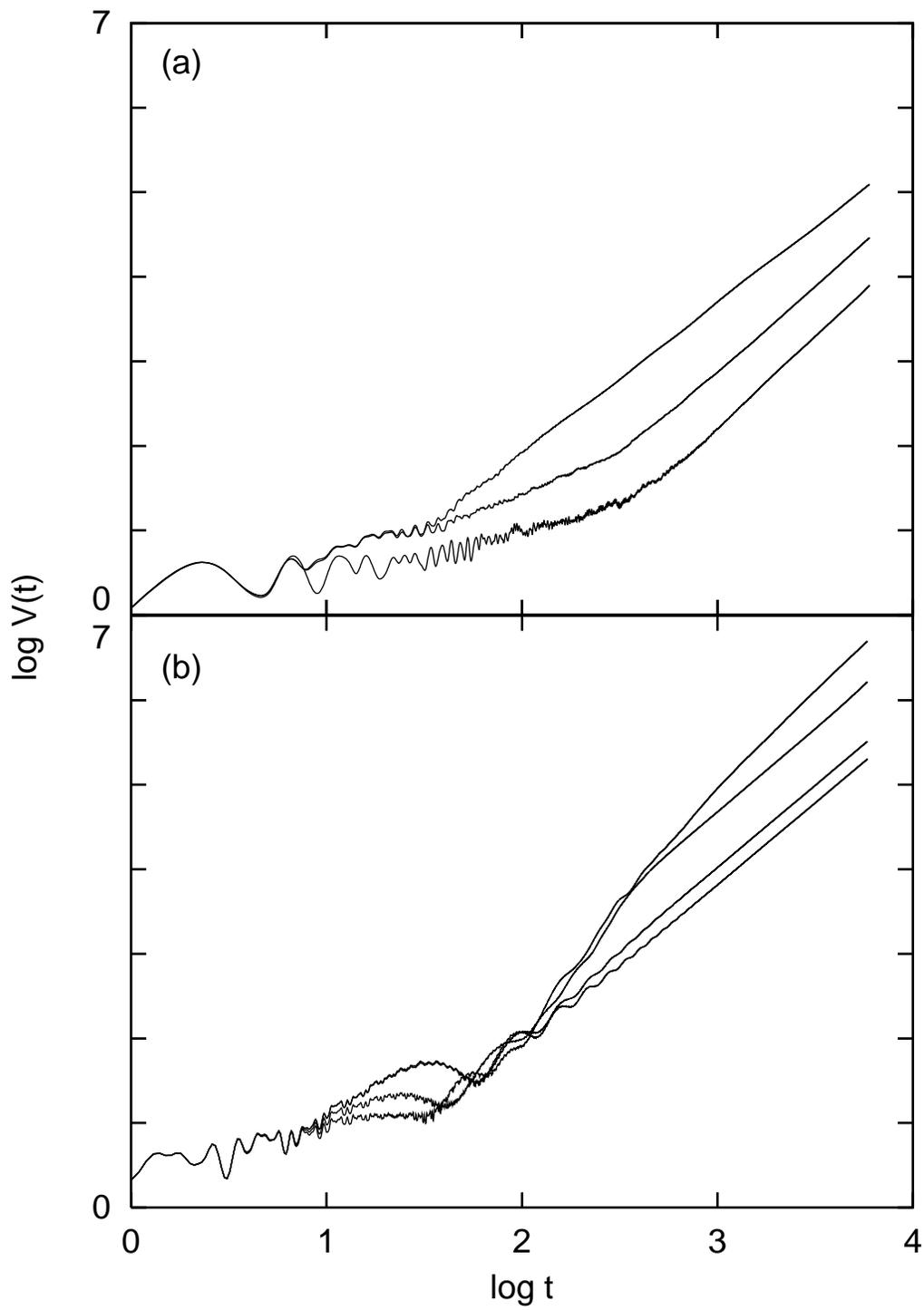
\noindent{\bf Fig.~8}.
$\log_{10}V(t)$ versus $\log_{10}t$ within the transfer model: (a)
$T=5.0$ with $\alpha=0, 0.2$, and $0.8$, and (b) $T=0.2$ with
$\alpha=0.6, 0.4, 0.1$, and $0$ from top to bottom at $t=1000$.

\end{multicols}
\end{document}